\documentclass[conference]{IEEEtran}

\usepackage{graphicx,mathptmx,amsmath,amsfonts,amsthm,color,comment}
\usepackage{enumitem}
\usepackage{enumerate}
\usepackage{algorithmic}
\usepackage{pgfplots}
\usepackage{pgf}
\usepackage{tikz}

\def\R{\mathbb{R}}
\def\HN{ H }
\def\Hmin{ H_{\min} }

\def\LambdaN{ \Lambda^N }
\def\IN{ I^N }
\def\Noise{ N }

\def\Hmin{ H_{\min} }
\def\Cmin{ C_{\min} }

\begin{document}
\title{Physical layer insecurity}
\author{
\IEEEauthorblockN{Muriel M\'edard}
\IEEEauthorblockA{\textit{Research Laboratory for Electronics} \\
\textit{Massachusetts Institute of Technology}\\
Cambridge, USA \\
medard@mit.edu}
\and
\IEEEauthorblockN{Ken R. Duffy}
\IEEEauthorblockA{\textit{Hamilton Institute} \\
\textit{Maynooth University}\\
Maynooth, Ireland \\
ken.duffy@mu.ie}
}

\date{October 2022}
\maketitle
\begin{abstract}
In the classic wiretap model, Alice wishes to reliably communicate to Bob without being overheard by Eve who is eavesdropping over a degraded channel. Systems for achieving that physical layer security often rely on an error correction code whose rate is below the Shannon capacity of Alice and Bob's channel, so Bob can reliably decode, but above Alice and Eve's, so Eve cannot reliably decode. For the finite block length regime, several metrics have been proposed to characterise information leakage. Here we reassess a metric, the success exponent, and demonstrate it can be operationalized through the use of Guessing Random Additive Noise Decoding (GRAND) to compromise the physical-layer security of any moderate length code. 

Success exponents are the natural beyond-capacity analogue of error exponents that characterise the probability that a maximum likelihood decoding is correct when the code-rate is above Shannon capacity, which is exponentially decaying in the code-length. In the finite blocklength regime, success exponents can be used to approximately evaluate the frequency with which Eve's decoding is correct in beyond-capacity channel conditions. Through the use of GRAND, we demonstrate that Eve can constrain her decoding procedure through a query-number threshold so that when she does identify a decoding, it is correct with high likelihood, significantly compromising Alice and Bob's communication by truthfully revealing a proportion of it. 

We provide general mathematical expressions for the determination of success exponents in channels that can have temporally correlated noise as well as for the evaluation of Eve's query number threshold, using the binary symmetric channel as a worked example. As GRAND algorithms are code-book agnostic and can decode any code structure, we provide empirical results for Random Linear Codes as exemplars. Simulation results mimic the mathematical predictions, demonstrating the practical possibility of compromising physical layer security.

\end{abstract}

\section{Introduction}
Since Wyner's classic considerations \cite{Wyn75, OW84, CH77, CK78}, there has been a rich literature on the topic of wiretap channels and associated codes \cite{BB11, ZWB18}. In particular, much of the work on physical layer security has relied on the premise of exploiting the difference in signal to noise ratio between the sender, Alice, and the intended receiver, Bob, and the pair formed by Alice and the eavesdropper, Eve. In the limit of long codes, the additional noise that Eve experiences can be transformed into an effect that acts like a partial one-time pad, obstructing Eve's attempts to decode. The premise behind operational proposals is the design of codes \cite{TDCMM07, MV11, Demetal11} that are decodable by Bob under lighter noise but not by Eve under heavier noise. 

A natural question concerns the case when codes are not in the infinite limit setting and so concentration to the mean is no longer achieved with high probability, with short wiretap code constructions, of lengths of order 16, have gained attention \cite{ND17,nooraiepour2020secure,RC22}.

As Bob seeks to have reliable communication with Alice, the relevant concern is to limit the cases where he decodes in error. For such events, there is a wealth of techniques to bound their probability, for example error exponents \cite{Gal68, Wyn67, MO77, KSU10, BF02, Gal73}. For shorter codes, dispersion approximations and other considerations have been proposed \cite{PPV10, MGUP16, KK18}.  

For wiretap settings, both error exponents \cite{BTM17, Hay06, MM11, PTM16, Hay11}, as well as dispersion bounds and related finite blocklength analysis techniques \cite{Xuetal22, YSP19, Har20, HB19, Pfietal17, FAB16, SH21, Zheetal20, SLC21, Caoetal15, ATM22, Hay06} have been considered in the context of eavesdroppers. From the point of view of Eve, however, the security consideration is altogether different from reliable communication. Concern lies in the events where Eve is able to decode because the noise realization is advantageous as any successful decoding represents a failure of security \cite{merhav2014exact}. 

We show in this paper that the success probability of Eve is an informative metric. While transmissions above capacity cannot all be decoded reliably, it does not mean that they are all decoded with probability near zero. Moreover, for a range of rates above capacity, we demonstrate that Eve can identify decodings that she is confident are not in error, compromising the security of Alice and Bob's communication. 

In the context of developing Guessing Random Additive Noise Decoding (GRAND),
Proposition 1 of \cite{duffy19GRAND} establishes success exponents, where the probability that a maximum likelihood decoder successfully decodes given the code-rate is above capacity decays exponentially in block-length \cite{Ari73,dueck1979reliability}, for channels whose noise can have temporal correlation. These can be used to generate an estimate of the likelihood that Eve can correctly decode beyond capacity. Theorem 3 of \cite{duffy19GRAND} gives an additional result that provides Eve with a simple criterion to test whether she is confident in the proposed decoding. That is, not only can Eve decode correctly a fraction of the time, she can identify which decodings are likely to be correct.

In this paper, we explore Eve's ability to confidently and correctly decode moderate length codes acting beyond Shannon capacity when she uses GRAND. We consider random linear codes (RLCs), which achieve secrecy capacity in the case where the channel from Alice to Bob and the channel from Alice to Eve are both BSCs \cite{CV20} and  which have been previously considered for use in the wiretap channel setting \cite{BTM17}.

We note that while Bob generally seeks to be efficient in order to have a timely decoding for a well operating communications channel, Eve has no such constraint generally, so her guessing can be extensive. Our results show, however, that by using GRAND, Eve computational effort is minimal. Indeed, even though her signal to noise ratio is worse than Bob's, she can decode with a non-vanishing probability, which is sufficiently high to compromise the security of the wiretap system, while performing on average no more work that Bob, who cannot limit himself to decoding only in the most favorable noise realizations.

\section{Success exponents - theory}
\label{sec:theory}

Consider a simple version of the wiretap channel \cite{Wyn75, VH79} where Alice has binary data that she wishes to send Bob. Eve has an independent, noisier channel than Bob's. For both channels, we assume that the noise effect is independent of the transmitted code-word and additive but need not be independent and identically distributed. Both Bob and Eve have a statistical characterisation of their channels and perform hard detection decoding of Alice's error correcting code.

In this setting, Gallager's \cite{gallager1973random} error exponent results state that if a random, or random linear, code is used with a code-rate $R$ that is less than the Shannon capacity of the channel, $C=1-H$, where $H$ is the Shannon entropy rate of the noise, then the likelihood that a ML decoding is in error decreases exponentially in $n$ at a speed that depends on $R$ and $C$. The mirroring concept is success exponents, where the probability that a ML decoder successfully decodes given $R>1-H$ decays exponentially in $n$.

In particular, assume Alice transmits the coded information bits $x^n$ and Eve receives a version corrupted by  not-necessarily i.i.d. binary noise, $y^n = x^n \oplus \Noise^n$, where $\oplus$ is modulo $2$ summation. Eve performs GRAND decoding by sequentially taking, in order from most likely to least likely based on the noise statistics, putative noise effects, $z^n$, removing them from the received sequence and querying if what remains, $y^n \ominus z^n$, is in the code-book. The first instance where a code-book element is found is an ML decoding. 

For code-rates less than capacity, Gallager's error exponent can be extracted from GRAND by analysing the likelihood that the number of queries until the true noise effect, $N^n$, is encountered is greater than the number of queries that would identify a non-transmitted code-word. For code-rates greater than capacity, success exponents can be extracted by analysing the likelihood that the true noise effect is encountered while querying before the first noise effect that would result in finding a non-transmitted codeword. 

With all logs being base 2, to characterise the success exponent, define the R\'enyi entropy rate of the noise
$\{\Noise^n\}$ process with parameter $\alpha\in(0,1)\cup(1,\infty)$ to be
\begin{align*}
H_\alpha 
        = \lim_{n\to\infty} \frac 1n \frac{1}{1-\alpha}\log\left(\sum_{z^n\in\{0,1\}^n} P(\Noise^n=z^n)^\alpha\right),
\end{align*}
with $\HN=H_1$ being the Shannon entropy rate of the noise. 
Denote the min-entropy rate of the noise by $\Hmin =
\lim_{\alpha\to\infty} H_\alpha$. 
Using these definitions of $H_\alpha$ it has been established \cite{Arikan96,Malone04,Pfister04,Christiansen13} that the moments of the distribution of the number of queries required to identify a realization of the noise effect, $\Noise^n$, when questions are asked in decreasing order of likelihood, scale exponentially with a rate that can be identified in terms of the R\'enyi entropy rates
\begin{align*}
\LambdaN(\alpha)
=
\begin{cases}  
        \displaystyle\alpha  H_{1/(1+\alpha)}
        & \text { for } \alpha\in(-1,\infty)\\
        -\Hmin & \text{ for } \alpha\leq-1.
\end{cases}
\end{align*}
Define the Legendre-Fenchel transform of $\LambdaN$ to be 
\begin{align*}
    \IN(g) = \sup_{\alpha\in\R} \left\{g\alpha -\LambdaN(\alpha)\right\},
\end{align*}
which is the rate function for a large deviation principle of the scaled logarithm of guesswork \cite{Ari2000,Christiansen13}. Proposition 1 of \cite{duffy19GRAND} proves that if $R>C$, the probability then an ML decoder provides a correct decoding decays exponentially in $n$ with rate $\IN(1-R)$.

Theorem 3 of \cite{duffy19GRAND} gives a conditional result that if $0 < g < 1 - R$ is such that $\IN(g) < 1-R-g$, then the probability of a correct decoding given GRAND made fewer than $2^{ng}$ queries converges to $1$ as $n$ increases. If the code rate $R > C$, then a sufficient condition for there to be an exponent for number of queries below which concentration to a correct result occurs is that $R<1-\Hmin = \Cmin$, where we will call $\Cmin$ the min-capacity. Taken together, these results establish that to ensure that, when equipped with a hard detection decoder, Eve can never confidently decode any of the communication between Alice and Bob, it is necessary that $R>\Cmin$ rather than $R>C$.

\begin{figure*}[htbp]
\centerline{\includegraphics[width=0.9\textwidth]{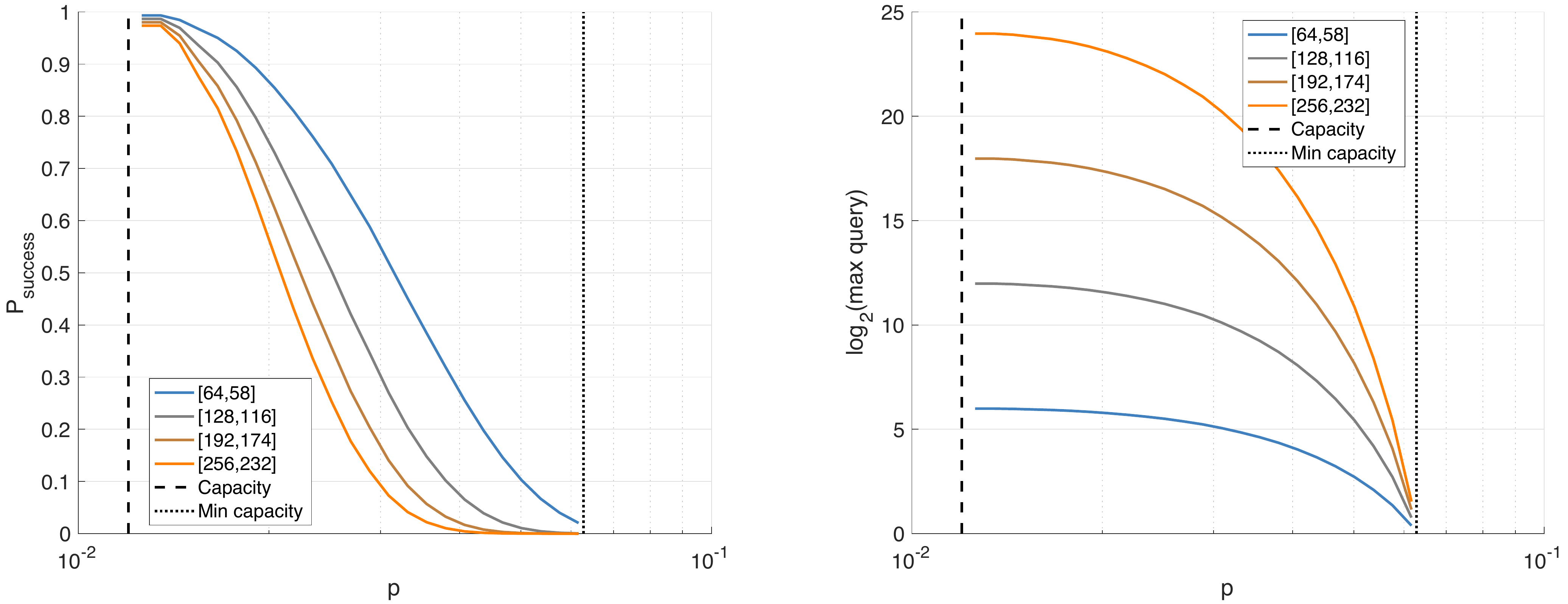}}
\caption{For a BSC and four codes of different lengths, but the same rate rate of approximately $0.91$, figures are plotted for channel conditions between Shannon capacity and min-capacity. Left hand panel: the approximate probability of a successful ML decoding as determined from the success exponent, $\exp(-n \IN(1-R))$, for the given BSC $p$. Right hand panel, theoretical predictions of $\log_2$ of the maximum query number that Eve should perform with GRAND to retain confidence in a correct decoding. Empirical results for random linear codes are presented later.}
\label{fig:Beyond_Capacity_Theory}
\end{figure*}

As a worked example, assume that the channel between Alice and Eve is a BSC with bit-flip probability $p$. In that case
\begin{align*}
\LambdaN(\alpha) = \begin{cases} 
        (1+\alpha)\log\left((1-p)^{\frac{1}{1+\alpha}} + p^{\frac{1}{1+\alpha}}\right) & \text{ if } \alpha\in(-1,\infty)\\
        \log(\max(1-p,p)) & \text{ if } \alpha \leq-1,
        \end{cases}
\end{align*}
and both the success exponent $\IN(1-R)$ and the exponent for the maximum number of code-book queries that would result in a confident decoding can be readily evaluated numerically. 

Using those formulae to create finite block length approximations for codes of rate $\approx 0.91$ and lengths $n=64, 128, 192$ and $256$, results in Fig. \ref{fig:Beyond_Capacity_Theory}. The left hand panel shows the approximate likelihood that an ML decoding would be correct. The right hand panel shows $\log_2$ of the mathematically determined approximate maximum number of queries that can be made while ensuring that correct decoding is highly likely to be returned. These predictions can be compared with the empirical results that follow.

\section{Success Probabilities - Practice}

\begin{figure*}[htbp]
\centerline{\includegraphics[width=0.9\textwidth]{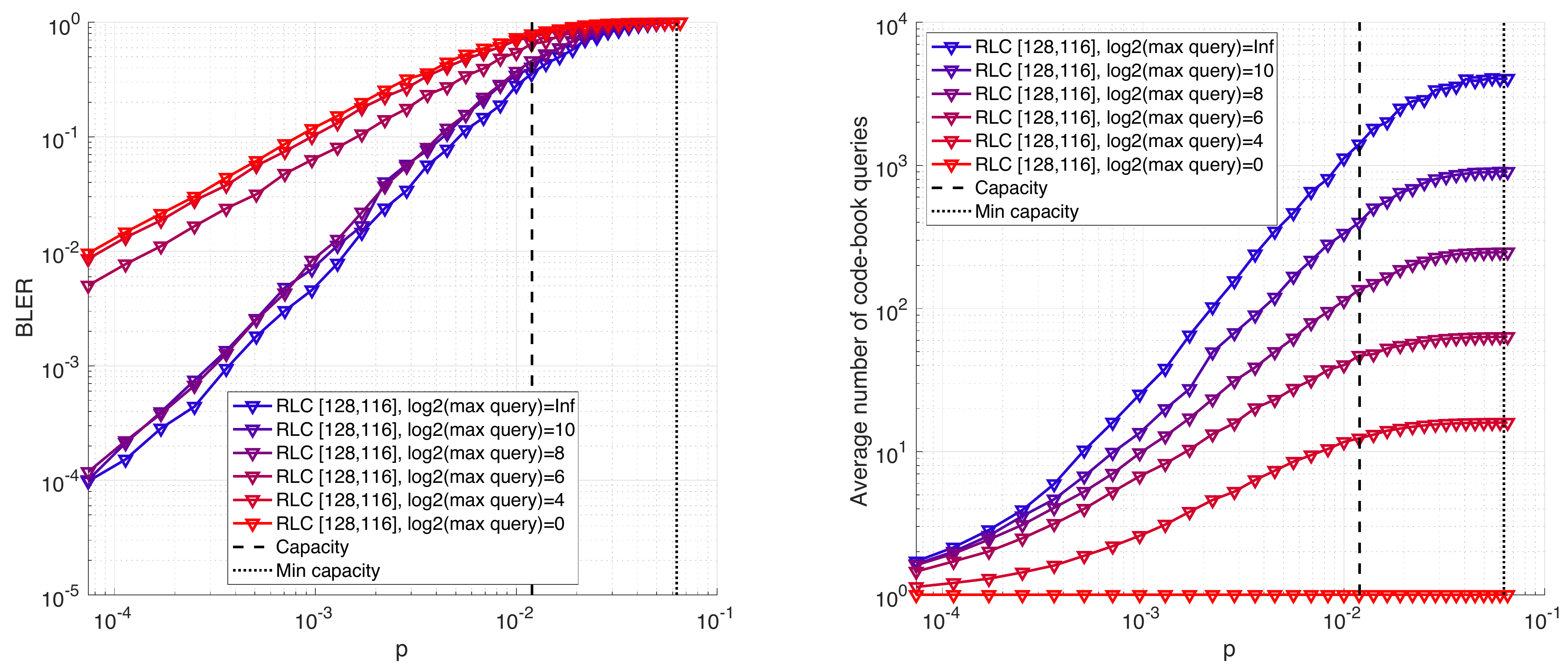}}
\caption{A random linear [128, 116] code transmitted over a BSC with bit flip probability $p$ decoded with GRAND. Left hand panel: block error rate. Right hand panel: average number of queries until a decoding is found or abandonment occurs. The blue line results from running hard detection GRAND-BSC without abandonment. As the colour gets redder, GRAND is abandoning earlier, at $2^a$ queries for the $a$ stated in the legend, which is recorded as an error. }
\label{fig:Beyond_Capacity_RLC_128_116}
\end{figure*}

We first consider the natural analogue of the setting for the results  in Fig. \ref{fig:Beyond_Capacity_Theory} where Alice and Bob are using a random linear $[128,116]$ code. Eve decodes transmissions using GRAND and abandons after $2^a$ queries, where $a$ is reported in figure legends, returning an erasure rather than erroneous decoding, allowing her to discard  decodings she would not trust.

Fig. \ref{fig:Beyond_Capacity_RLC_128_116} focuses on $R<C$ by showing Block Error Rate (BLER) against $-\log_{10}$ of the BSC's bit flip probability where abandonment counts as an incorrect decoding. As Theorem 2 of \cite{duffy19GRAND} establishes that all GRAND algorithms will identify an incorrect decoding after approximately $2^{n(1-R)}=2^{n-k}$ queries, abandoning a little before then does not impact BLER performance when $R<C$ as decodings that would likely be in error are instead erasures. If abandonment happens too early, the full within-capacity decoding performance of the code is not realised. The right panel shows the average number of queries until a code-word is found or abandonment, where abandoning early can be seen to significantly reduce complexity. 

\begin{figure*}[htbp]
\centerline{\includegraphics[width=0.9\textwidth]{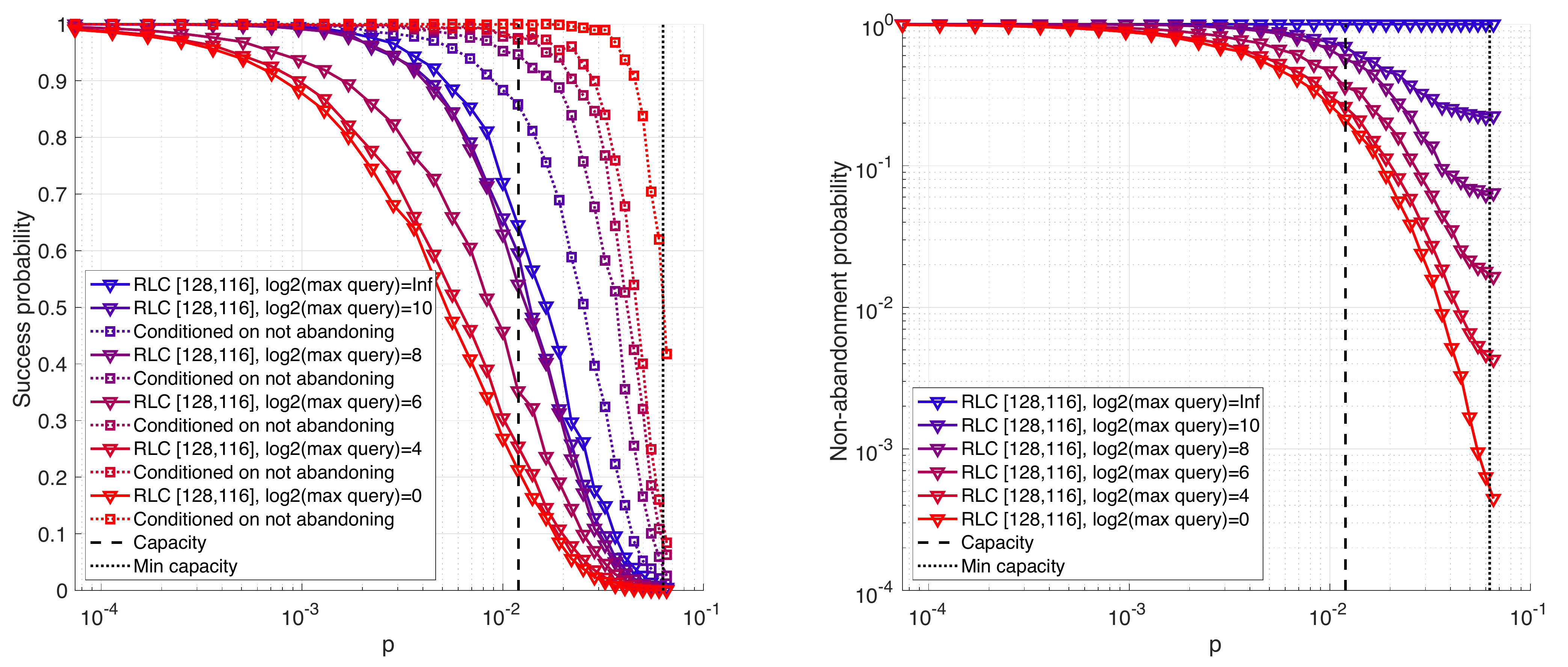}}
\caption{Same setup as Fig. \ref{fig:Beyond_Capacity_RLC_128_116}. Left hand panel: shows the likelihood that the decoding is correct, the success probability, where abandonment is counted as incorrect. The dotted lines are the success probability conditioned on not abandoning. Right hand panel: the proportion of non-abandoned decodings. The dashed black vertical line marks the capacity threshold, where $R<C$ to the left and $R>C$ to the right. The dotted black vertical line marks the min-capacity threshold, where $R<\Cmin$ to the left and $R>\Cmin$ to the right.}
\label{fig:Beyond_Capacity_RLC_128_116_success}
\end{figure*}

The left panel of Fig. \ref{fig:Beyond_Capacity_RLC_128_116_success} replots the data in the first panel of Fig. \ref{fig:Beyond_Capacity_RLC_128_116} but focuses on the region where the code-rate is above capacity, $R>C$. It shows the likelihood that the decoding is correct, the success probability, where abandonment is counted as incorrect, and can be compared to the left hand panel of Fig. \ref{fig:Beyond_Capacity_Theory}. This speaks to the success exponent for all ML decoders showing graceful degradation as code-rate passes through capacity. 

The dashed lines in Fig. \ref{fig:Beyond_Capacity_RLC_128_116_success} are the success probability conditioned on not abandoning. By abandoning decoding early, the conditional success probability can be kept high for values of $p$ well above Shannon capacity and up to min-capacity. The right panel of the figure shows the proportion of non-abandoned decodings, which decreases as the abandonment threshold is reduced, and is reflective of the fraction of Alice to Bob communications that Eve decodes rather than abandons. 

\begin{figure*}[htbp]
\centerline{\includegraphics[width=0.9\textwidth]{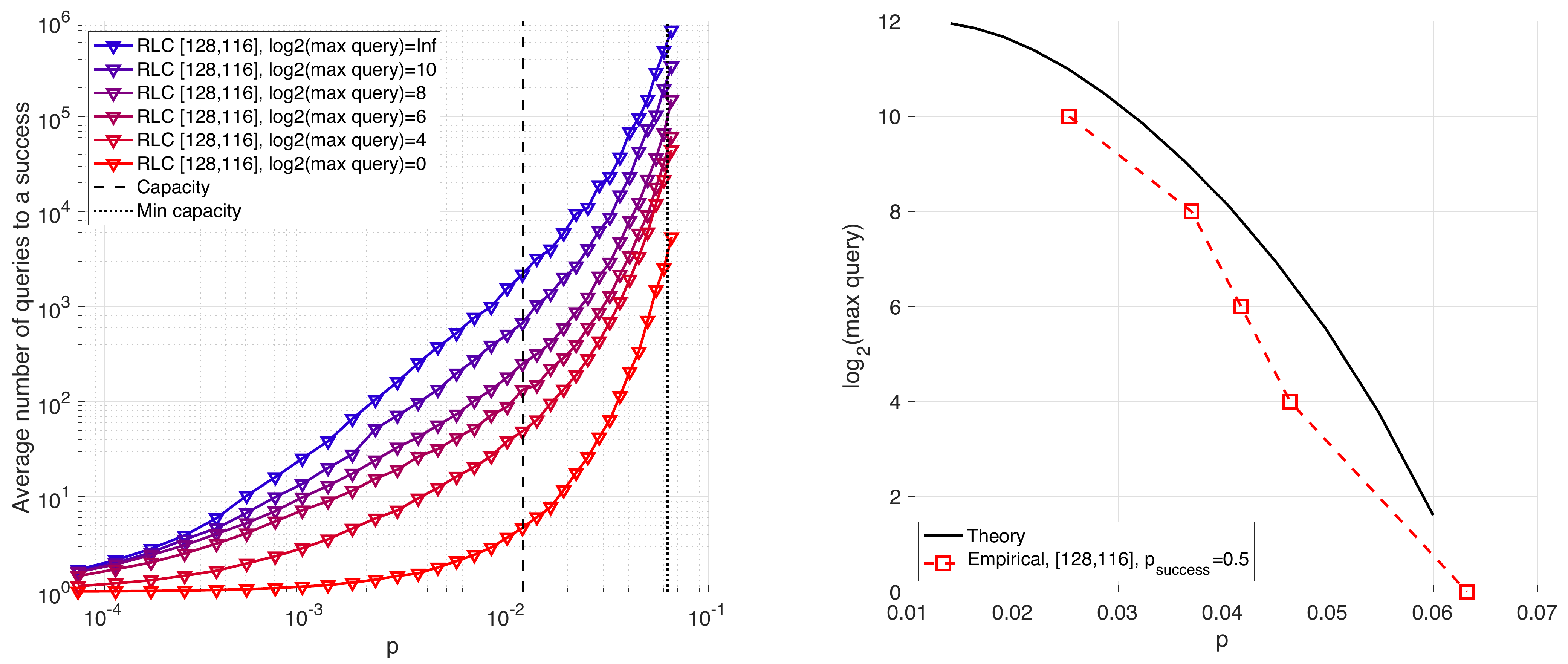}}
\caption{Same setup as Fig. \ref{fig:Beyond_Capacity_RLC_128_116}. Left hand panel: shows the average number of queries. including those that lead to an abandonment until a correct decoding. Right hand panel: black line is the theoretically determined number of queries that should be allowable while still concentrating on a correct decoding. The red line is the empirically determined value for which 50\% of decodings given non-abandonment are correct for the abandonment criterion in the y-axis.}
\label{fig:Beyond_Capacity_RLC_128_116_theory}
\end{figure*}

Fig. \ref{fig:Beyond_Capacity_RLC_128_116_theory}, left panel, provides an indicator of how little work Eve is doing between non-abandoned decodings in terms of the total number of queries until she identifies a decoding she is confident in, where the early abandonment limits the effort used in decoding what would be untrustworthy decodings anyway. The right panel of Fig. \ref{fig:Beyond_Capacity_RLC_128_116_theory} plots the mathematical determination of $\log_2$ of the maximum number of queries that Eve should be able to make while ensuring she only returns confident decodings (black lines - as determined using the formulae in Sec. \ref{sec:theory}), while the red line identifies the empirical equivalent which ensures than more than 50\% of Eve's decodings are correct. This establishes that, given knowledge of the bit-flip probability, Eve can use theory to guide her abandonment threshold. The observations from these results are that by only trusting decodings that are identified before a thresholded number of queries, when Eve does decode she can be confident that the decoding is correct, so long as $R<\Cmin$, compromising Alice and Bob's communication.

\begin{figure*}[htbp]
\centerline{\includegraphics[width=0.9\textwidth]{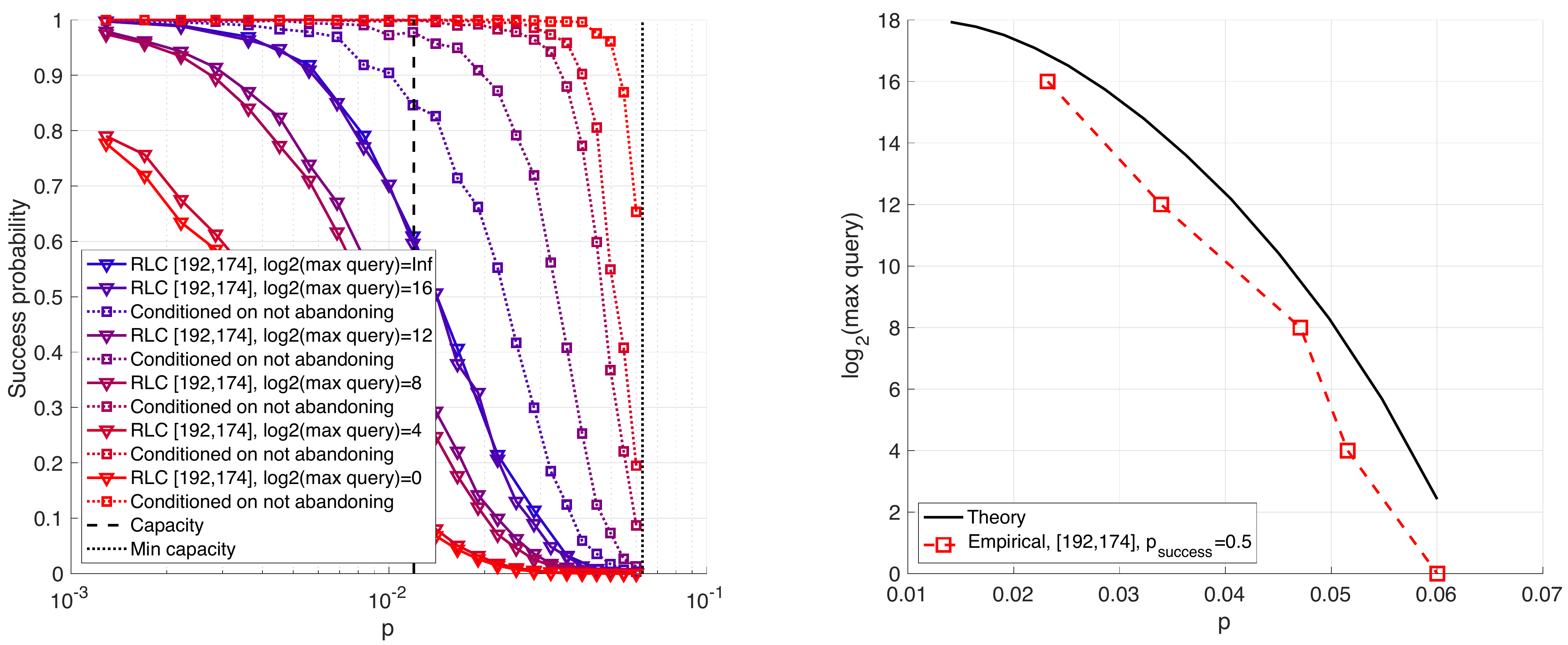}}
\caption{Similar setup to Fig. \ref{fig:Beyond_Capacity_RLC_128_116_success}, but for a [192,174] RLC. Left hand panel: the success probability, where abandonment is counted as incorrect. The dotted lines are the success probability conditioned on not abandoning. The dashed black vertical line marks the capacity threshold, where $R<C$ to the left and $R>C$ to the right. The dotted black vertical line marks the min-capacity threshold, where $R<\Cmin$ to the left and $R>\Cmin$ to the right.
Right hand panel: black line is the theoretically determined number of queries that should be allowable while still concentrating on a correct decoding. The red line is the empirically determined value for which 50\% of decodings given non-abandonment are correct for the abandonment criterion in the y-axis.}
\label{fig:Beyond_Capacity_RLC_192_174_success_theory}
\end{figure*}

For a longer [192,174] RLC of the same rate as the [128,116] code, the left hand panel of Fig. \ref{fig:Beyond_Capacity_RLC_192_174_success_theory} shows analogous results to the left panel of Fig. \ref{fig:Beyond_Capacity_RLC_128_116_success} on the success probability and conditional success probability. Despite being a longer code that has greater total redundancy, the behavior is similar and consistent with theoretical predictions. The right hand panel is akin to that of Fig. \ref{fig:Beyond_Capacity_RLC_128_116_theory}, demonstrating the correspondence between the theoretical approximate evaluation of an appropriate abandonment threshold and the empirically identified one, where again good correspondence is observed.

\section{Discussion}

For codes whose rate are above Shannon capacity, we have demonstrated how success exponents can be used to estimate the fraction of correct decodings returned by a maximum likelihood decoder. We have illustrated how an eavesdropper, Eve, can use an abandonment threshold with GRAND, which can operate with any code, to identify decodings that are confidently correct to compromise Alice and Bob's communication. 

For the BSC example used here, Eve's abandonment threshold can be related to a Hamming weight of the corresponding noise effect, but the mathematics and method can also be used for channels with memory where that would not be the case. Moreover, as GRAND algorithms search for the noise-effect, a BSC is essentially the worst case hard detection setting as the noise has minimal structure. In practice, channels are not memoryless but rendered synthetically so through interleaving, which is commonly part of the construction of wiretap systems \cite{Aja21, PF22, Haretal10, FAB19, XJL18}. Noise correlation, however, increases the effective SNR, possibly by multiple dBs, and has been demonstrated to be exploitable in the GRAND framework to improve decoding accuracy \cite{duffy19GRAND, An22}. Even if Bob uses a scheme that relies on interleaving, Eve can use statistical knowledge of the correlation of the noise, resulting in an effectively higher SNR, i.e. lesser degradation, of her channel relative to Bob's, while also using the mathematical formulation to inform her abandonment thresholds.

We have focused here on the simple case of a BSC, which is representative of DMCs, to which the approach can be applied. Hard-detection hardware implementations of GRAND for BSC and bursty channels have been published \cite{Riaz21, Riaz22,abbas2020,abbas2021high-MO} that demonstrate it is an efficient decoding algorithm for moderate redundancy codes. Fading channels can be more vulnerable than channels without fading \cite{BR06}. 
In the setting where the channel noise is represented in a continuous domain, e.g. as Gaussian, the use of soft information, such as through ORBGRAND \cite{duffy2021ordered, duffy22ORBGRAND}, which is close to capacity achieving \cite{liu2022orbgrand,Yuan22} and practically implementable in hardware \cite{abbas2022high, condo2022fixed, Riaz23}, would further aid Eve in compromising Alice and Bob's communication.

\bibliographystyle{IEEEtran}
\bibliography{grand,ORBGRAND}

\end{document}